\documentclass[12pt,oneside,peerreview,draftcls,onecolumn]{IEEEtran}

\usepackage{amsmath,amsfonts,amssymb,amsbsy, amsthm,ae,aecompl}
\usepackage{algorithm,algpseudocode}
\usepackage[utf8]{inputenc}
\usepackage[english]{babel}
\usepackage{bm}
\usepackage{color}
\usepackage{cite}
\usepackage{epsfig}
\usepackage{enumerate}
\usepackage{float}
\usepackage{fancyhdr}
\usepackage[T1]{fontenc}
\usepackage[acronym,toc,shortcuts]{glossaries}
\usepackage{graphicx, caption, subcaption}
\usepackage{hyperref}
\usepackage{bbm}
\usepackage{lastpage}
\usepackage{listings}
\usepackage{lipsum}
\usepackage{dsfont}
\usepackage{overpic}
\usepackage{transparent}
\usepackage{tablefootnote}
\usepackage{tikz,pgfplots}
\usepackage{times}
\usepackage{verbatim}
\usepackage[all]{xy}

\hypersetup{
    colorlinks,%
    citecolor=black,%
    filecolor=black,%
    linkcolor=black,%
    urlcolor=black
}

\graphicspath{{figures/}}

\bgroup
\makeatletter
\renewcommand{\@seccntformat}[1]{%
  \ifcsname prefix@#1\endcsname
    \csname prefix@#1\endcsname
  \else
    \csname the#1\endcsname\quad
  \fi}
\newcommand\prefix@section{\thesection. }
\makeatother

\makeglossaries
\newacronym{API}{API}{application programming interface}
\newacronym{BS}{BS}{base station}
\newacronym{massive-MIMO}{massive-MIMO}{massive multiple-input multiple-output}
\newacronym{MBS}{MBS}{macro base station}
\newacronym{iid}{i.i.d.}{independent and identically distributed}
\newacronym{PPP}{PPP}{{P}oisson point process}
\newacronym{SBS}{SBS}{small base station}
\newacronym{SCN}{SCN}{small cell network}
\newacronym{SIR}{SIR}{signal-to-interference ratio}
\newacronym{SINR}{SINR}{Signal-to-Interference-plus-Noise Ratio}
\newacronym{UT}{UT}{user terminal}
\newacronym{QoS}{QoS}{quality-of-service}
\newacronym{QoE}{QoE}{quality-of-experience}
\newacronym{GUI}{GUI}{graphical user interface}

\begin{document}  
\title{Wireless Caching: Technical Misconceptions and Business Barriers}
\author{
\IEEEauthorblockN{
Georgios Paschos$^{\diamond}$, 
Ejder Baştuğ$^{\star}$, 
Ingmar Land$^{\diamond}$, 
Giuseppe Caire$^{\dagger}$, and  
Mérouane Debbah$^{\diamond}$}\\
\IEEEauthorblockA{\small 	\vspace{0.2cm}
$^{\diamond}$Huawei Technologies, France Research Center, Mathematical and Algorithmic Sciences Lab,\\
	20 Quai du Point du Jour, Boulogne-Billancourt, 92100, France\\
$^{\star}$Large Networks and Systems Group (LANEAS), CentraleSupélec, Université Paris-Saclay,\\
	3 rue Joliot-Curie,  91192 Gif-sur-Yvette, France \\
$^{\dagger}$Technische Universit{\"a}t Berlin, Electrical Engineering and Computer Science \\
Einsteinufer 25, 10587, Berlin, Germany\\
\{georgios.paschos, ingmar.land, merouane.debbah\}@huawei.com \\
ejder.bastug@centralesupelec.fr, 
giuseppe.caire@tu-berlin.de \\ \vspace{-0.9cm}
}
\thanks{This research has been partly supported by the ERC Starting Grant 305123 MORE (Advanced Mathematical Tools for Complex Network Engineering), and the project BESTCOM.}
}
\IEEEoverridecommandlockouts
\maketitle
\IEEEpeerreviewmaketitle

\begin{abstract}
Caching is a hot research topic and poised to  develop into a key technology for the upcoming 5G wireless networks.  The successful implementation of caching techniques however, crucially depends on joint  research developments in different scientific domains such as networking, information theory, machine learning, and wireless communications. Moreover, there exist business barriers related to the complex interactions between the involved stakeholders, the users, the cellular operators, and the Internet content providers.
In this article we discuss  several technical misconceptions with the aim to uncover enabling research directions for   caching in  wireless systems. Ultimately we make a  speculative stakeholder analysis for wireless caching in 5G.
\end{abstract}
\begin{IEEEkeywords}
Edge caching, 5G, Wireless networks.
\end{IEEEkeywords}

\section{Introduction}
\label{sec:introduction}
\emph{Caching is a mature idea from the domains of web caching, content delivery networks, and memory optimization in operating systems. Why is caching still an active topic of discussion?} 
In the 90s, the traffic in the web exploded, leading its inventor Sir Tim Berners-Lee to declare the network congestion  as one of the main challenges for the Internet of the future. The congestion was caused by the \emph{dotcom boom} and specifically due to the client-server model of connectivity, whereby a webpage was downloaded from the same network server by every Internet user in the world. The challenge was ultimately resolved by the invention of Content Delivery Networks (CDNs), and the exploitation of web caching. The latter replicates  popular content in many geographical areas and saves bandwidth by avoiding unnecessary multihop retransmissions. As a byproduct, it also  decreases access time (latency) by decreasing the distance between the two communicating entities.

Today, 30 years later, we are reviving the same challenge in the wireless domain. The latest report of Cisco \cite{Cisco2015} predicts a massive increase of Internet devices connected through the wireless access, and warns of 
a steep increase in mobile traffic which is expected to reach by 2018 roughly 60\% of total network traffic, the majority of which will be video. The wireless  system designers strive to fortify 5G wireless networks with higher access rates on the one hand and with increased densification of network infrastructure on the other. Over the last three decades,  these two  approaches are responsible for the majority of  network capacity upgrade per unit area,  successfully absorbing   the wireless traffic growth. However, with the explosion of access rates and number of  base stations, the backhaul of wireless networks will also become congested \cite{Bastug2014LivingOnTheEdge, Wang2014Cache} which motivates further the use of caching: 
 store popular reusable information at the base stations to reduce the load at the   backhaul.
Furthermore a recent technique \cite{MaddahAli2014Fundamental} combined caching with coding and revolutionized how goodput scales in bandwidth-limited networks. 
Therefore,  caching has the potential to become the third key technology for wireless systems sustainability. 

\begin{figure*}[ht!]
	\centering
	\includegraphics[width=0.95\linewidth]{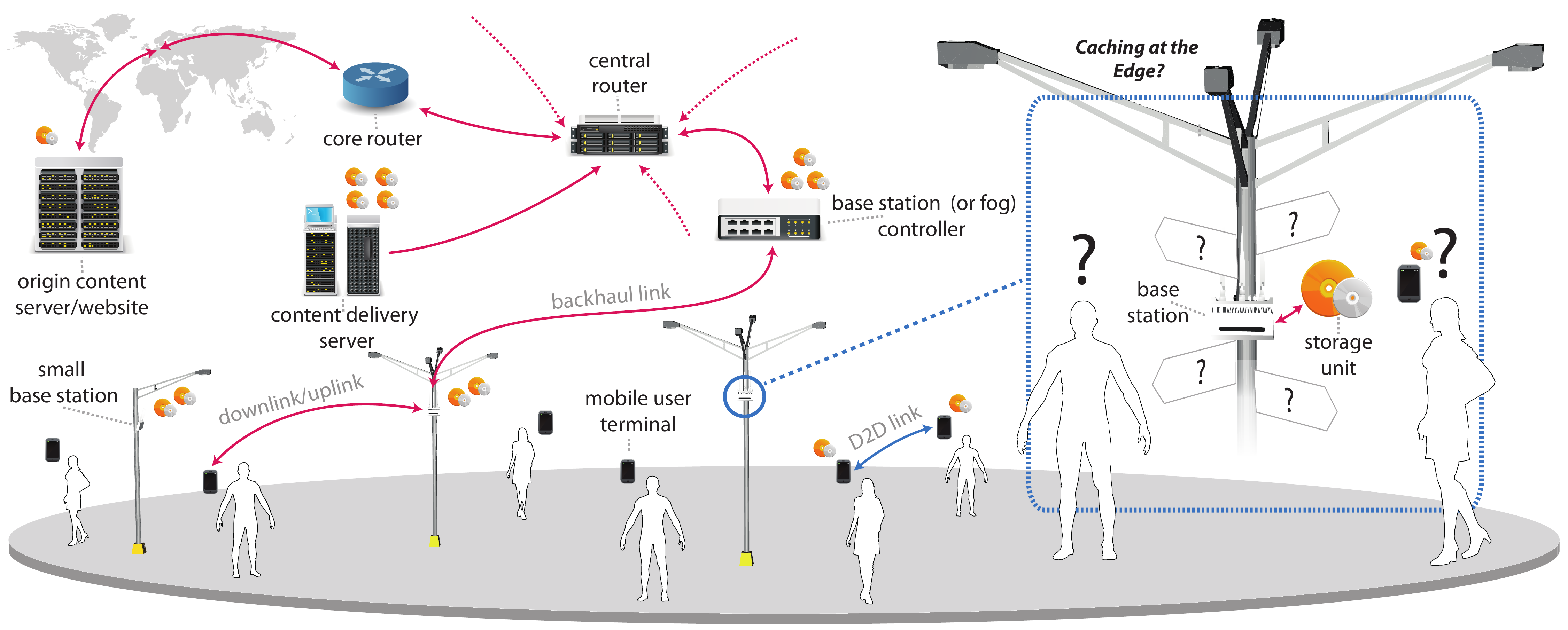}
	\caption{An illustration of caching in future wireless networks. Contents available in the origin server are cached at the base stations and user devices, for offloading the backhaul and the wireless links.}
	\label{fig:scenario}
\end{figure*}

The research community is converging to an enabling architecture as the one described in Figure~\ref{fig:scenario}. In the  network of the future, memory units can be installed in gateway routers between the wireless network and the Internet (e.g. in 4G this is called S-GW), in base stations of different sizes (small or regular size cells), and in end-user devices (e.g. mobile phones, laptops, routers, etc). 
In this article, we  discuss important topics such as (a) the characteristics of cacheable content and how this affects caching technologies in wireless, (b) where to install memory, and (c) the differences between wireless caching and legacy caching techniques. Last, we focus on business barriers that must be overcome for the successful adoption of wireless caching by the industry.

\section{Dealing with Massive Content}
Not all  network traffic is cacheable. Interactive applications, gaming, voice calls, and remote control signals are examples of information objects that are not reusable and hence cannot be cached. Nevertheless, most network traffic today (estimated 60\% \cite{Cisco2015}) is deemed cacheable. We refer to cacheable information objects as \emph{content} in this article. Since the performance of caching is inherently connected to the specifics of contents, this section is dedicated to the understanding  of these specifics.
In particular we focus on the following misconceptions: (i) The static IRM model is sufficient for experimentation, (ii) user information cannot be used for popularity estimation due to the vast number of users, and (iii) security issues precludes caching at the edge.
%
\subsection{Insufficiency of Static Popularity Models}
The standard approach to design and analyze caching systems involves a model for generating content requests to replace the actual request traces--this  approach is  often several  orders of magnitude faster.  
The de facto model for performance analysis of web caching is the Independence Reference Model (IRM):   content $n$ is requested according to an  independent Poisson process with rate $\lambda p_n$, where $p_n$ refers to the content popularity, modelled by a power law, i.e.,  $p_n\propto n^{-\alpha}, ~\alpha>0$. This well-established model thrives due to its simplicity; it only has two parameters, namely  $\lambda$ to control the rate of requests, and $\alpha$ to control the skewness of the popularity. 

Numerous studies  fit IRM to real traffic with satisfactory results \cite{Zipf}, so why do we need to change it? 
The IRM  assumes that the content popularity is static, which is of course not true. Trending tweets, breaking news, and the next episode of Game of Thrones, are examples of ephemeral content with rapidly changing popularity; they appear, they become increasingly popular, and they gradually become unpopular again.  

\begin{figure}[ht!]
	\centering
	\includegraphics[width=0.65\linewidth]{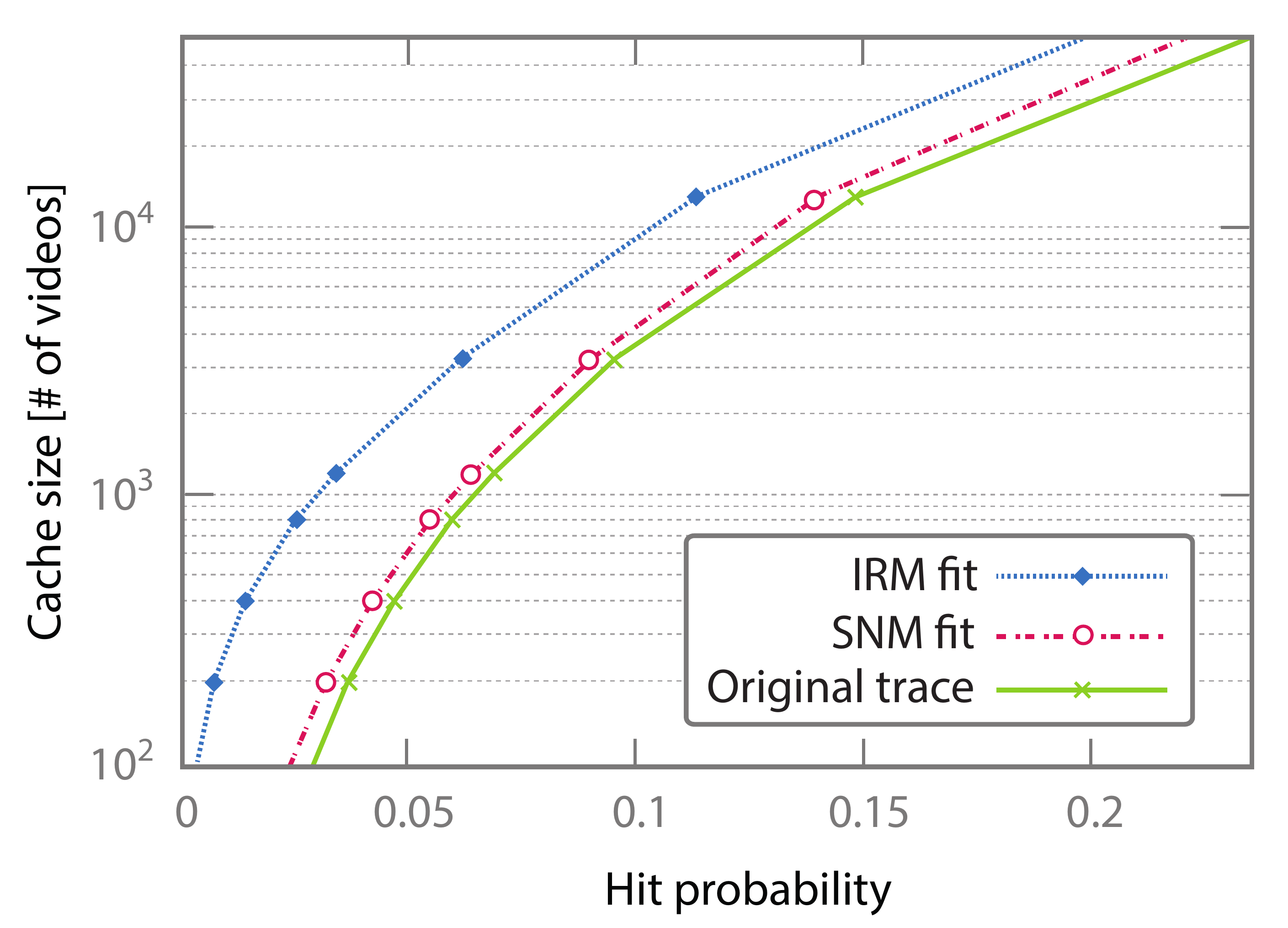}
	\caption{(from \cite{leonardi}) Hit probability comparison between best fit of IRM, SNM  and the YouTube  traces.}
	\label{fig:snm}
\end{figure}

In fact, \cite{leonardi}  considers large YouTube and Video on Demand (VoD) datasets and discovers:  time-varying models are more accurate than IRM with respect to caching performance analysis--Figure~\ref{fig:snm} reproduces the comparison when fitting YouTube data and shows the superiority of modeling the popularity as time-varying. In the  inhomogeneous Poisson model proposed in \cite{leonardi} each content is associated to a ``pulse'' whose duration reflects  the content lifespan and whose height denotes its instantaneous popularity--the model is called the \emph{Shot Noise Model} (SNM), mirroring the Poisson noise from electronics. 
 While the shape of the pulse is not important, the study  observes strong correlations between popularity and duration; apparently popular contents prosper longer. Finally, a class-based model  \cite{leonardi} can  conveniently capture spatio-temporal correlations while allowing analytical tractability.
 Mobile users are especially keen on  downloading  ephemeral content, thus it is expected that in case  of  wireless content the improvement in modeling accuracy will be even greater.  
 
To optimize a cache one needs to track the changes in content popularity. 
For example, the classical  web caching  systems  adopt  dynamic eviction policies like Least-Recently-Used (LRU) in order to  combat time-varying content popularity in a heuristic manner.
However, the joint consideration of popularity variations with wireless systems reveals a new challenge that renders LRU policies inefficient. 
While a typical CDN cache normally receives 50 requests/content/day,  the corresponding figure for  base station cache may be as low as 0.1 requests/content/day. With such a small number of requests, fast variations of popularity become very difficult to be tracked and classical LRU schemes fail. 
 
This development motivates novel caching techniques that employ learning methodologies to accurately track  the evolution of content popularity over time. A recent study \cite{mathieu} analyzes the SNM model and gives the optimal\footnote{The optimality is established for the restricted case of homogeneous rectangularly-shaped pulses and asymptotically large content catalogs.} policy for joint caching and popularity estimation. Additionally \cite{Roberts2} proposes as an alternative solution the use of LRU with prefilters. 
\subsection{How to Track Popularity Variations}
Since content popularity is time-varying, the caching operations can only be optimized if a fresh view of the system is  maintained. This requires massive data collection and processing, and statistical inference from this data, which by itself is a complex task to handle. Additionally, user privacy is a concern that can limit the potential of collecting such information. So can we timely gather all this information in a wireless network? 

Consider $K$ users subscribed to the telecom operator  and  $L$ caches  placed in the network (e.g., in base stations or other locations). Each of these caches has capability of storing $M$ contents out of $N$ contents in the catalog. Then, let the matrix ${\mathbf P} \in \mathbb{R}^{K \times N}$ model the content access statistics where rows are users and columns are contents. In other words, each entry (or rating) in this matrix quantifies how popular is content $n$ to user $k$.  The popularity matrix ${\mathbf P}$ is large, sparse, and only partially known in practice, and has to be continuously estimated in order to enable the correct cache decisions at the base stations. 
At first, this seems to be an impossible feat.

To deal with the complexity of handling matrix ${\mathbf P}$, it is possible to use  machine learning tools to estimate the unknown entries. Such an estimation is particularly efficient when the matrix has low spectral dimension, ~and the system can be described by a small number of ``features''; fortunately popularity correlation induces such behaviour in  matrices obtained from real data. For instance,  low-rank matrix factorization methods, i.e., ${\mathbf P} \approx  {\bf K}^T {\bf N}$ where ${\bf K} \in \mathbb{R}^{r \times K}$ and ${\bf N} \in \mathbb{R}^{r \times N}$ are factor matrices, can be employed to construct the $r$-rank version of the matrix, using the fact that users' interests are correlated and predictable when $r$ is small. This additionally allows to store the collected statistics in a  more compact way.
As a result, a big data platform installed in the operator network can provide efficient  collection and processing of  user access patterns from several locations, as evidenced in \cite{Bastug2015BigData}. 
Further development of novel machine learning tools, such as clustering techniques, are needed to  improve the estimation of the time-evolving content popularity matrix, i.e.,   ${\mathbf P}_l(t)$ for base station $l$, which may  differ from base station to base station. 

It is worth noting that a caching system has  requirements similar to those  of a  recommendation system. For example,  the well-known Netflix movie recommendation system exploits information of user's past activity in order to predict which movie is likely to be scored high by the user. Similarly, a caching system exploits the request sequence  to predict what contents are popular enough to be cached. In this context, user privacy  regulations may affect the collection of these valuable data.
 
A key topic of research in this direction is privacy-preserving mechanisms that can enable sufficient sampling of the time-evolving and location dependent popularity matrix ${\mathbf P}_l(t)$ without compromising user privacy. Moreover, it is interesting to consider learning enhancement approaches that transfer information to caching domain from other domains such as "transfer learning" \cite{Bastug2015TransferExtended}.

\subsection{Security Is A Kind of Death}
A common anti-caching argument relates to the operation of caching in a secure environment. 
The secure counterpart of HTTP protocol, called HTTPS, was originally  used to provide end-to-end (e2e) encryption  for securing sensitive information like online banking transactions and authentication. Owing to the recent  adoption from traffic giants Netflix and YouTube, the HTTPS protocol is growing in numbers to soon exceed  50\% of the total network traffic. Content encryption poses an unsurmountable obstacle to in-network operations, including caching. 
Since encrypting  the data makes them unique and not reusable, caching,  or even statistically processing encrypted content is impossible. Ironically, the statement ``security is a kind of death'' of Tennessee Williams seems to squarely apply to wireless caching.

Security is definitely a precious good everyone welcomes. Although securing a video stream might seem as an excessive measure, in some cases it may be well justified. Unfortunately, e2e encryption is clearly not in the Berners-Lee spirit since it  prevents operators from optimizing their networks and reanimates the server-client ghost of congestion, a reality that equally  no one can overlook.  In fact,  modern CDN systems resolve this issue by having ``representatives'' of the content provider at the edge of the Internet. These representatives are trusted entities which hold the user keys and are able to decrypt the requests and perform standard caching operations. Ultimately, this methodology is neither entirely secure for the user \cite{Carnavalet2016Killed}, nor efficient for the network \cite{Maisonneuve2015Security, papagiannaki}. The need to make the system sustainable finally overrules the need for e2e encryption, which is an argument against HTTPS for video delivery. Given however this situation, how can we realistically push caching deeper into the wireless access?

Currently, content providers install their own caching boxes in the operator network and intercept the related encrypted content requests deeper in the wireless access network. Examples include the Google Global Cache \cite{GoogleGlobalCache}, the Saguna solution \cite{Saguna}, and the CacheBOX of Appliansys \cite{CacheBox}. In this approach, the boxes are not controlled by the operator which leads to several limitations:  (a) The caching boxes cannot perform complex tasks. (b) It is difficult to apply  learning techniques without   context information from the operator. (c) The caching approach is similar to the CDNs and therefore does not exploit the performance opportunities specific to wireless caching, as we will discuss below.

New security protocols have been proposed to  enable the operators to perform caching on encrypted requests \cite{papagiannaki}.  This leads to an interesting research direction, to  combine the user security and privacy with  facilitation of the network  management operations, which are crucial for the sustainability of the future wireless systems.

\section{Towards a Unified Network Memory}
The proposition of Information Centric Networking (ICN) as a candidate for the future Internet has also risen the subject of \emph{where to install network memory} \cite{Roberts}. The ICN approach proposed to equip routers with caches, and to allow content replication everywhere in the network. 
A recent work \cite{Shenker}  came up with striking conclusions about the ICN approach: most of the caching benefits of ICN can be obtained by caching at the edges of the network using the existing CDNs, and that any extra caching  in the core network brings only negligible improvements at very high costs. In the wireless domain, however, the question remains relevant: does it make sense to cache even closer to the user than CDN? 

It is commonly believed that  caching is very inefficient near the user, and thus it should be done at CDN. Below we explain the main reasons of inefficiency and argue that they can be overcome.
%
\subsection{Caching Deeper than CDN}
Mitigating backhaul and wireless link overload requires going beyond CDN and caching at the base stations and the mobile users. 
However, the efficient operation of such caches is very  challenging. In particular, there are two main challenges: (a)  caches used in wireless networks are typically small as compared to CDN caches, and (b) the popularity profile of traffic is highly unpredictable when non-aggregated.

To understand point (a), consider that the effectiveness of a cache is measured with the \emph{hit probability}, i.e., the fraction of requests found in the cache. This can be  upper bounded by the  popularity sum $\sum_{n=1}^Mp_n$, where $p_n$ is  the ordered popularity distribution with $p_1$ denoting the probability to request the most popular file. 
For power-law popularity the sum can further be approximated by $(M/N)^{1-\alpha}$, where $\alpha<1$ is the power-law exponent. A very small ratio $M/N$ means that the  hit probability becomes vanishingly small. For example, if we are caching Netflix (12.5PB) in a mobile phone (10GB), $M/N\sim 10^{-6}$, $\alpha\sim 0.8$ and the hit probability is  less than $10\%$. However, base stations equipped with a disk array (40TB) can be extremely effective when caching contents for a mobile VoD application. Table \ref{fig:table} provides some indicative numbers for the memory types available and the catalogue sizes of reasonable applications.

In this context there are three promising research directions: (i) choose a part of the catalog to cache while maintaining network neutrality \cite{Kocak2013network}, (ii) store only parts of the content using partial caching techniques \cite{maggi}, and (iii) install massive memory at the edge in the form of small-sized datacenters \cite{Bioglio2015Optimizing}. The third option will be realized by the \emph{fog computing} paradigm \cite{Bonomi2012Fog}.

\begin{table}
\centering
    \begin{tabular}{|lc|c|c|c|}
         \hline
            \multicolumn{2}{|l|}{ }  & {\bf Netflix catalogue} ($12.5$PB)\tablefootnote{The entire catalogue was anecdotally measured to contain $3.14$PB of content in 2013, which however we multiply by $4$ since the same video is available in multiple formats.}  & {\bf Torrents} ($1.5$PB)   & {\bf Wireless VoD catalogue} ($1$TB)   \\ \hline \hline
             {\bf Disk}   		& $2$TB   	& $\sim 0.01\%$  & $\sim0.1\%$ & $100\%$ \\ \hline
             {\bf Disk Array}    & $40$TB  	& $\sim 0.3\%$    & $\sim 2\%$  & $100\%$ \\ \hline
             {\bf Data Center} & $150$PB   & $50\%$     & $100\%$ & $100\%$ \\
         \hline
    \end{tabular}
    \caption{Typical data size values for normalized cache size $M/N$ taken from the study of \cite{Roberts2}. In practice it is anticipated that the wireless traffic is a 80-20 mix of torrent-like traffic and live VoD traffic tailored to wireless device capabilities.}
    \label{fig:table}
\end{table}

\begin{figure*}[ht!]
	\centering
	\includegraphics[width=0.95\linewidth]{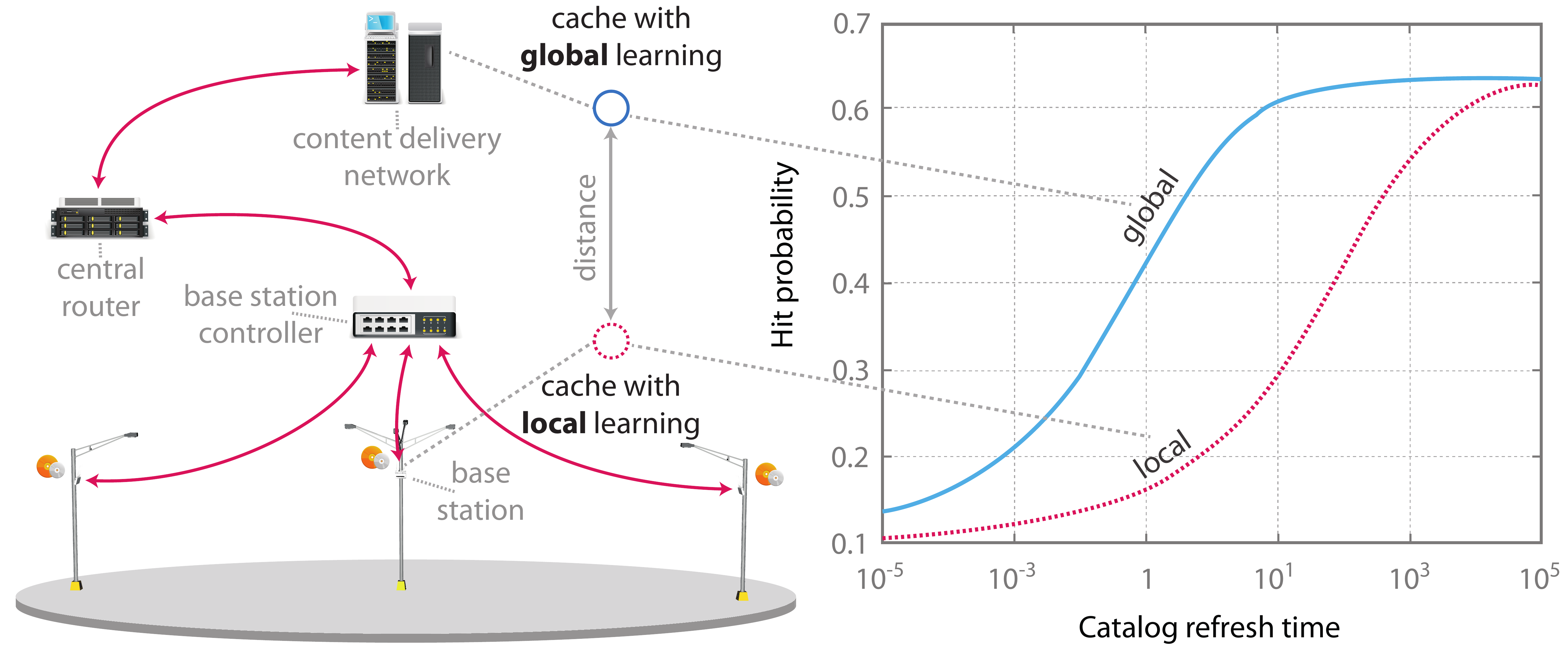}
	\caption{Optimal hit probability comparison between observing the aggregate request process at the CDN-level (global), and observing the individual request process at each base station cache (local), when  refreshing the catalogue. The hit probability performance depends on how fast the time-varying popularities can be learnt: global is faster than local.}
	\label{fig:global}
\end{figure*}

To understand the unpredictable nature of sparse requests (formulated as challenge (b) above), consider as an example the delivery of breaking e-news in a city served by a single CDN node. Most users will download the news only once. The CDN system can quickly detect the rising popularity of the news, since it will receive many requests in a short time frame. From the point of view of a mobile user, however, the detection of the popularity of the trending news becomes very difficult because the news is requested only once by the given user. This example shows that the detection efficiency    depends on the number of requests  aggregated at the popularity learner. To illustrate this, Figure~\ref{fig:global} shows the optimal hit probability  in a hierarchy of $L$ base stations.   Learning at the global CDN cache is shown to detect variations that are $L$ times faster than those at local caches. To remedy the situation it is possible to use an architecture that combines  information obtained at different aggregation layers  \cite{mathieu}.
%
\subsection{Memory Is Cheap But Not Free}
Although the cost of a small cache is dwarfed by  the base station cost,  the total amount of  installed memory in a mobile network can be  considerable, therefore deciding to install wireless caching requires a careful cost analysis \cite{Roberts2}. To compute the optimal  size of memory to install at each location, one needs to know (a)  the cost coefficients, (b)  the skewness of content popularity, and (c) the local traffic distribution in cells. Predicting how  (a)-(b) will evolve is quite challenging, but as in \cite{Roberts2} a survey may determine a good set of parameters at any given time.

\begin{figure*}[ht!]
	\centering
	\includegraphics[width=0.95\linewidth]{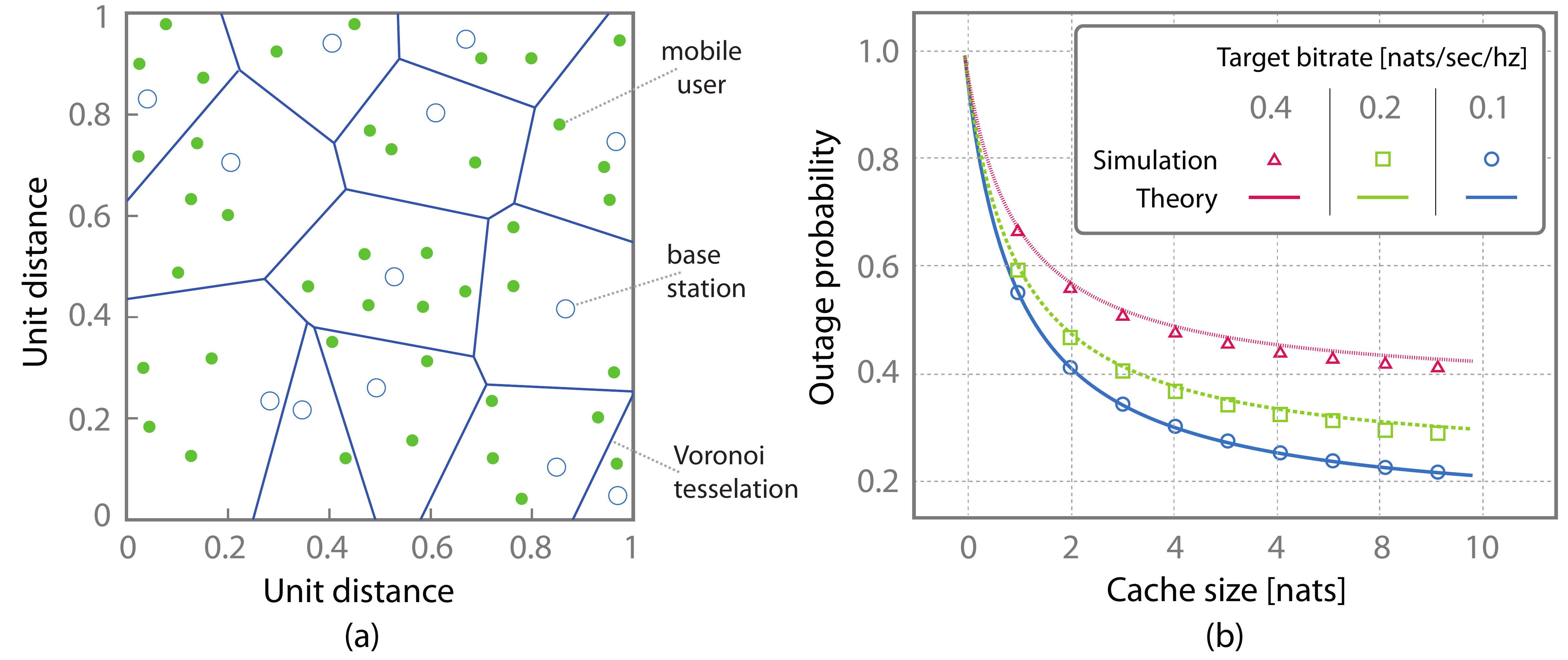}
	\caption{An example of base station deployment with caching capabilities \cite{Bastug2014CacheEnabledExtended}: (a) illustrates a snapshot of 2D topology in which users and cache-enabled base stations are distributed according to Poisson Point Processes (PPPs), (b) shows theoretical performance gains of such a  deployment, where the validation of results is done via simulations.}
	\label{fig:ppp}
\end{figure*}

For (c), the literature is extensively based on grid models, which  in the case of future wireless networks might be off for a significant factor. More accurate models have been recently introduced from the field of  stochastic geometry, where the cache-enabled base stations are distributed according to a spatial point process (often chosen to be 2D  Poisson), thus enabling to handle the problem analytically. The validity of such modelling compared to regular cellular models has been verified using extensive simulations. 
Additional insights for the deployment of cache-enabled base stations can be obtained by analytically characterizing the performance metrics, such as the outage probability and average delivery rate, for a given set of parameters such as given number of base stations, storage size, skewness of the distribution, transmit power and target \ac{SINR}  \cite{Bastug2014CacheEnabledExtended}.  In addition to initially studied single-tier network \cite{Bastug2014CacheEnabledExtended}, the more detailed modeling and analysis of heterogeneous networks \cite{Yang2015Analysis, Chen2016Cooperative}, online caching policies \cite{Zaidi2015Information}, uniform caching \cite{Blaszczyszyn2014Geographic}, power consumption  \cite{Perabathini2015CachingGreen}, and a Markovian mobility \cite{Poularakis2013Exploiting}  are some recent examples in  this direction. Therefore, although  storage units become increasingly cheaper, the question of how much storage we should place at each location should be studied jointly with  topological models from stochastic geometry, which are both tractable and realistic. An example for such modelling and analysis of a network with cache-enabled base stations is provided in Figure \ref{fig:ppp}. 
\section{Wireless $\neq$ Wired}
Web caching has been traditionally studied by the  networking community. A common misconception says that caching is a network layer technique, and hence the web caching approaches  are sufficient for  wireless caching as well.

However, following the fundamental work of Maddah-Ali and Niesen  \cite{MaddahAli2014Fundamental}, the idea of caching has penetrated  the information theory community with a new twist termed \emph{coded caching}, which promises unprecedented gains. In the following, we discuss the differences between wired and wireless caching.
%
\subsection{Wireless Caching Lies Both at Network and PHY Layers}
Suppose that a base station  wants to deliver information to $K$ users at a rate of 1Mbps each, i.e., for  streaming a  video. If the video is the same for all users (broadcast video) then this might be possible for an arbitrarily large number of users. For example the base station could use an omni-directional antenna, exploit the broadcast characteristic of wireless medium, and transmit at 1Mbps to all users simultaneously.   When the videos are different, this is clearly not possible: the base station needs to multiplex the users over frequency, time, or codes, where each such \emph{resource block} is then associated to a single user. Since the resource blocks are finitely many, ultimately the base station can serve 1Mbps videos up to a maximum number of users $K_{\max}$, after which the resources are exhausted. To increase $K_{\max}$, physical layer researchers propose ways to increase the resource blocks of a given spectrum, i.e., increase spectral efficiency, or install more base stations so there are more resource blocks per unit area, referred to as network densification.  

\begin{figure}[ht!]
 	\centering
	\includegraphics[width=0.65\linewidth]{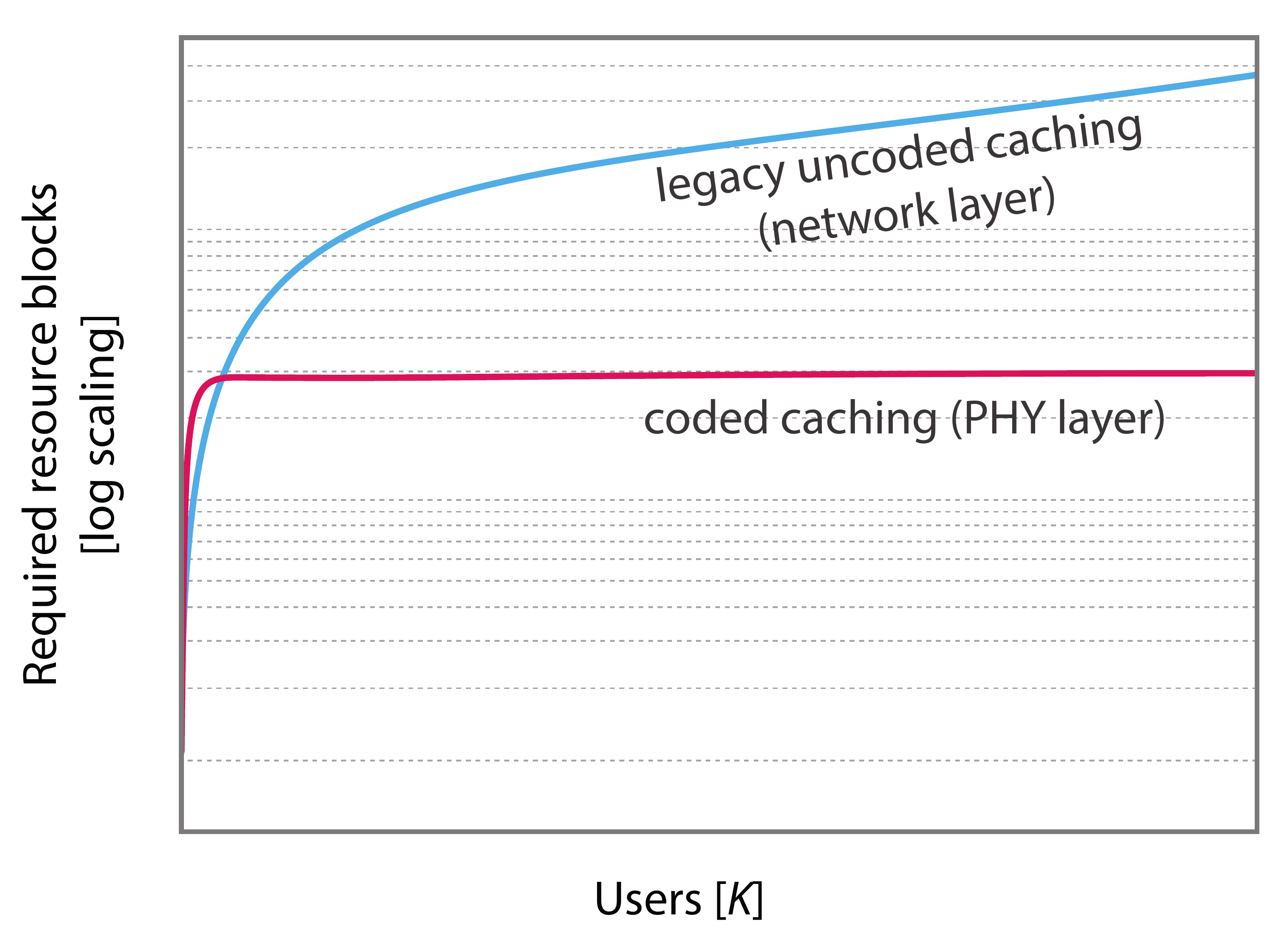}
	\caption{Required resource blocks for $K$ mobile users with unicast demands, when the caches fit $30\%$ of the catalog. 
	Coded caching can serve  an arbitrarily large population of users with a fixed number of resource blocks.  }
	\label{fig:scaling}
\end{figure}

The novel  paradigm  of \cite{MaddahAli2014Fundamental} shows a surprising fact:  exploiting caching in a smart way, an unbounded number of users watching different videos can be accomodated. 
  How is this made possible? During off-peak operation of the network, the users can cheaply populate their caches with \emph{parts} of popular contents. This is a perfectly reasonable assumption since the question of sustainability that caching is trying to tackle refers to  the hours of the day that the  network experiences the traffic peak.  The content parts are appropriately chosen according to a caching code which ensures  symmetric properties. Then at request time, a coding technique called ``index coding'' is employed, to minimize the number of transmissions to satisfy all users.\footnote{In fact, finding the optimal index code is a very difficult problem, and hence the proposed approach  resorts to efficient heuristics.} The combination of these schemes is shown in \cite{MaddahAli2014Fundamental} to yield required resource blocks equal to  
$K (1 - M/N) / (1 + K M/N ),$
where $K$ is the number of users, $M$ the cache size, and $N$ the catalog size.
Hence, if the cacheable fraction of the catalog $M/N$ is kept fixed, then the required
number of resource blocks does not increase with the number of users $K$, this can be verified by taking the limit $K\to\infty$ whereby the above quantity converges to a constant. The result is pictorially summarized in Figure~\ref{fig:scaling}.

This promising idea has sprung a wealth of research efforts, some of them in parallel or right after, such as device-to-device (D2D) networks \cite{Ji2014Fundamental, Jeon2015Capacity}, non-uniform content popularities \cite{Ji2015Order}, online caching policies \cite{Pedarsani2013Online}, multi-servers \cite{Shariatpanahi2015Multi}, multi-library \cite{Sahraei2016Multi} and  multihop wireless networks with cooperative caching \cite{Gitzenis2013Asymptotic}. More recently, it has been shown that in order to achieve such ``order of $K$'' gain over 
conventional unicast (with possible legacy uncoded caching) systems, the content objects must be split into an $O(\text{exp}(K))$ number of subpackets; for networks of practical size this gain is not achievable. The optimal tradeoff between coded caching gain and content object size is a very interesting topic of current research. 

From the implementation point of view, promising research directions include extensions to capture system aspects such as (i) popularity skewness, (ii) asynchronous requests, (iii) content objects of finite size and (iv) cache sizes that scale slower than $N$. Assuming that these practical challenges are resolved,  caching for wireless systems will become intertwined with physical layer techniques employed at the base station and the handheld.
%
\subsection{One Cache Analysis Is Not Sufficient}
A contemporary mobile receives the signal of more than 10 base stations simultaneously. In future densified cellular networks, the mobile will be connected to several femto-, pico-, or nano- cells. The phenomenon of wireless multi-access opens a new horizon in caching exploitation \cite{femtocaching}. Since a user can retrieve the requested content from many network endpoints, neighboring caches should  cooperate  and avoid storing the same objects multiple times.  

Content placement optimizations of wireless caching typically boil down to a  set cover  problem in a bipartite graph connecting the users to the reachable caches. Therefore, finding what contents to store at each cache is  a difficult problem  even if the popularities are assumed known \cite{femtocaching}. It is possible to relax the problem to convex optimization by the use of distributed storage codes, where each cache stores coded combinations of contents \cite{femtocaching}, or by obtaining a fractional placement by time sharing different integral placements. These ideas lead to several interesting algorithms in the literature of cooperative caching \cite{Poularakis2014Approximation, Naveen2015Interaction, Bioglio2015Optimizing, Zhang2015Fundamental, Dehghan2015Complexity}.

What is the gain from these approaches? Cooperative caching typically saves space in the cache by avoiding caching the same popular contents in neighboring caches. Equivalently we may think of  multiplying the cache size $M$ by a small number, at best say a gain of 3-5. With respect to hit probability, this can correspond to very different levels of gain, depending on the value of $M/N$. Due to the skewness of the popularity distribution, marginal hit probability gain\footnote{Gain obtained in hit probability  when increasing $M$ slightly.} is high when $M/N$ is small, and very small when $M/N$ is large. Since in  wireless we expect the former, high gains are expected from cooperative wireless caching.
 
The current proposals on cooperative caching assume static popularity, and therefore a promising direction of research along these lines is to design caching schemes that combine cooperation with learning of the time-varying popularity. The time to search and retrieve the content from a nearby cache may also be significant, hence intelligent hash-based filtering and routing schemes  are required \cite{Tao2015Content}. 
%
\section{A Stakeholder Analysis for Wireless Caching}
\label{sec:grand}
%
The business of wireless caching involves three key stakeholders that together form a complex ecosystem.

\textbf{The users} of telecommunication services are primarily the customers and the consumers of the content, but in the case of wireless caching they are also active stakeholders. The users might be requested to help in the form of contributing with their own resource, (for example in the case of coded caching it will be memory and processing, or in D2D caching it will be also relaying transmissions) and they will end up spending energy in the benefit of  better performance. On the other hand, one could envision the users  employing D2D technology to enable  caching without the participation of other stakeholders. Due to the complexities mentioned above however,  efficient wireless caching will require heavy coordination and extensive monitoring/processing, hence D2D approaches will be limited to restricted environments. 

\textbf{The operators} of telecommunication networks are well placed for wireless caching. Due to the particularities of coded caching and multi-access caching, the operators are in a unique position to implement new protocols in base stations, affect the standards for new mobile devices, and develop big data processing infrastructure that can realize wireless caching. Nevertheless, for reasons related to encryption, privacy, and global popularity estimation,  the operators might not be able to install these technologies without the cooperation of the other two stakeholders.

\textbf{The providers} of Internet content are champions of trust from the user community. Apart from the security keys, they also hold  extensive expertise in implementing caching techniques in the core networks. From this advantageous position, they can positively affect the progressive evolution of caching in wireless networks. On the other hand, content provider-only solutions cannot unleash the full potential of wireless caching, since they are limited to alienated boxes in the operator network that can perform caching only with legacy  CDN techniques. The deeper the caches go into the wireless network, the less efficient they will be if they stick to legacy CDN techniques. 

\begin{figure}[ht!]
	\centering
	\includegraphics[width=0.65\linewidth]{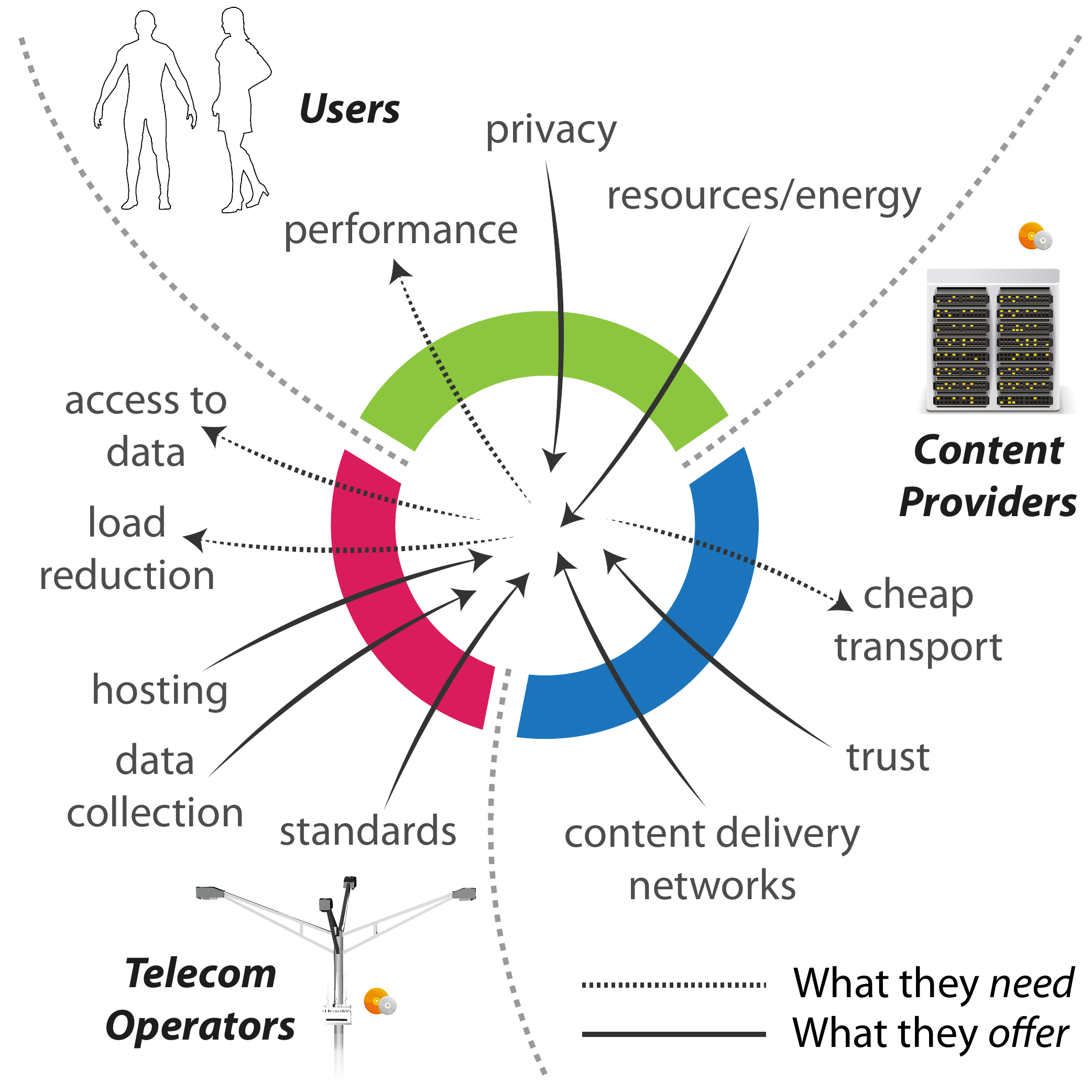}
	\caption{A stakeholder analysis.}
	\label{fig:stackeholders}
\end{figure} 

We summarize what each stakeholder \emph{offers} and \emph{needs} in Figure~\ref{fig:stackeholders}.
What are the prospects of the required collaboration among the stakeholders? Operators and content providers are seeking the ``best friends forever'' union \cite{Ft2012Content, Openet2014Content} in order to mutually harvest    benefits in the digital value chain while keeping their users happy. This is a favorable environment for the  scenario of wireless caching. 
In fact, if the telecom operators enable caching capabilities at the edge of their network,  their infrastructure will become sustainable while they will gain access to  new business models. On the other hand, content providers can benefit from a caching collaboration since (a)  traffic will be intercepted earlier and the content transport cost will be reduced, (b) the  user  demand will not be held back by sustainability issues, (c) costs associated to deployment of large memory units will be avoided, (d)  they will be able to reach closer to their users and extend computing infrastructures to the fog paradigm. Last, it is foreseeable that in some situations the role of the content provider and the wireless operator may  converge. 
\bibliographystyle{IEEEtran}
\bibliography{references}
\end{document}